\documentclass[useAMS,usenatbib]{mn2e}

\usepackage{color}
\usepackage{graphicx}
\title[Unstable mass-outflows in geometrically thick accretion flows
around black holes]
 { Unstable mass-outflows  in geometrically thick accretion flows
around black holes }
\author[Toru Okuda and Santabrata Das] {Toru Okuda$^{1}$
 \thanks{E-mail:bbnbh669@ybb.ne.jp}   and Santabrata Das$^{2}$
 \thanks{E-mai: sbdas@iitg.ernet.in } \\
$^{1}$ Nishi-Asahioka-Cho 3-15-1, Hakodate 042-0915, Hokkaido, Japan \\
$^{2}$ Indian Institute of Technology Guwahati, Guwahati, 781039, India}

\begin{document}

\date{Accepted }

\pagerange{\pageref{0}--\pageref{0}} \pubyear{2012}

\maketitle

\label{firstpage}

\begin{abstract}
 Accretion flows around black holes generally result in mass-outflows that exhibit irregular
 behavior quite often.
 Using 2D time-dependent hydrodynamical calculations, we show that
 the mass-outflow is unstable in the cases of thick accretion flows such as the low
 angular momentum accretion flow and the advection-dominated accretion flow.
 For the low angular momentum flow, the inward accreting matter on the equatorial plane
 interacts with the outflowing gas along the rotational axis and the centrifugally supported
 oblique shock is formed at the interface of both the flows, when
 the viscosity parameter $\alpha$ is as small as $\alpha \le 10^{-3}$.
 The hot and rarefied blobs, which result in the eruptive mass-outflow, are generated in
 the inner shocked
 region and grow up toward the outer boundary. The advection-dominated accretion flow attains
 finally in the form of a torus disc with the inner edge of the disc at  $3R_{\rm g} \le r \le 6R_{\rm g}$
 and the center at $6R_{\rm g} \le r \le 10R_{\rm g}$,
 and a series of hot blobs is intermittently formed
 near the inner edge of the torus and grows up along the outer surface of the torus.
 As a result, the luminosity and the mass-outflow rate are modulated irregularly where
 the luminosity is enhanced by $10-40\%$ and the mass-outflow rate is increaed by a factor of
 few up to ten.
 We interpret the unstable nature of the outflow to be due to the Kelvin-Helmholtz instability,
 examining the Richardson number for the Kelvin-Helmholtz criterion in the inner
 region of the flow.
 We propose that the flare phenomena of Sgr A* may be induced by the unstable mass-outflow as is
 found in this work.
\end{abstract}


\section{Introduction}

The observations as well as numerical simulations around black holes demosntrate regular,
irregular and/or chaotic time variation of emergent radiations and mass-outflows over a wide
range of time scales. The origins of such variable emissions and mass-outflows from the
vicinity of the black holes are also numerous. The accretion luminosity depends on the
adopted accretion models and its accretion rates. When the accretion rate is not too high, the
luminosity is directly proportional to the accretion rate and can be described successfully
by the standard thin-disc model (Shakura-Sunyaev, hereafter S-S, model) \citep{b30}.
The standard model have been proven to be successful in studying the cataclysmic
variables and later for neutron stars and black holes. However, for highly luminous accretion
discs whose luminosity exceeds the Eddington luminosity as is found in the black hole
candidate SS 433 or for very low luminous discs compared with the predicted accretion rate
as the case of the supermassive black hole candidate Sgr A*, the S-S model is not valid and
alternative thick disc models, such as the slim disc model \citep{b1}, the 
advection-dominated accretion disc model \citep{b22,b23} and the low angular momentum 
disc model \citep{b19,b8}, have been developed.

Numerous studies of the hot thick accretion flow around the black holes have been carried out
both theoretically and numerically to promote the insights of the black hole physics. 
One of the important results of the hot accretion flows is that the inflow and outflow rates in the accretion
flow follow a power-law function of radius. To explain the inflow-outflow rate versus radius  relation, 
two types of models of the adiabatic inflow-outflow solution \citep{b4,b5,b3} 
 and the convection-dominated accretion flow model \citep{b20,b29,b2,b14} have been proposed.
 Furthermore, the recent hydrodynamical (HD) and magneto-hydrodynamical (MHD) 
 simulations result in new findings of such outflows, winds and jets in the field of hot accretion flows around
 the black holes \citep{b21,b37,b43,b6,b15,b38}.
 The details of the geometrically thick hot accretion flows are referred to \citet{b36} and \citet{b40}.

 The hot accretion flows show unstable behaviors of the outflows in most cases of the time-dependent numerical simulations.
We have examined the thick accretion flows with low angular momentum and found that
the outflows from the outermost boundary are always unstable, accompanying irregularly eruptive mass ejections \citep{b25,b24}. 
 Although we interpreted the unstable mass-outflow 
to be induced by the irregular radial oscillations of the centrifugally supported shock,
the intrinsic origin of the unstable nature of mass-outflows is still an open question. 
In this paper,  focusing on the outflows from the hot accretion flow, we examine the unstable
 nature of the outflows in terms of the Kelvin-Helmholtz instability, using time-dependent 
2D hydrodynamical simulations. Finally, following the results of the unstable mass-outflow, 
we suggest a possible model which may explain the flaring events of Sgr A*.

\section{Numerical Methods}
\subsection{Basic equations}
  We consider a supermassive black hole with mass $M= 4\times 10^6 M_{\odot}$
  which is estimated for the supermassive black hole candidate  Sgr A*.
 We also assume a two-temperature plasma with ions being much hotter than electrons.
 The accreting gas is assumed to be very rarefied and optically thin.
  The set of relevant time-dependent equations is given in the spherical polar coordinates
  ($r$,$\zeta$,$\varphi$):

\begin{equation}
  { \partial\rho\over\partial t} + {\rm div}(\rho\bmath{v}) =  0,  
\end{equation}
\begin{eqnarray}   
 {\partial(\rho v)\over \partial t} +{\rm div}(\rho v \bmath{v}) & =&
  \rho\left[{w^2\over r}  +  {v_\varphi^2\over r} - {GM \over (r-R_{\rm g})^2} 
 \right] -{\partial p\over \partial r} \nonumber \\
  & &  +{\rm div}(\bmath{S}_r) + {1\over r}S_{rr},
\end{eqnarray}
\begin{equation}  
 {{\partial(\rho rw)}\over \partial t}+{\rm div}(\rho rw\bmath{v}) 
 = -\rho v_\varphi^2{\rm tan}\zeta-{\partial p\over\partial\zeta}
    + {\rm div}(r\bmath{S}_\zeta)
   + S_{\varphi\varphi}{\rm tan}\;\zeta, 
\end{equation}

\begin{equation}    
{{\partial(\rho r{\rm cos}\zeta v_\varphi)}\over \partial t} 
    +{\rm div}(\rho r{\rm cos}
\zeta v_\varphi\bmath{v}) =  {\rm div}(r{\rm cos}\;\zeta\bmath{S}_\varphi), 
\end{equation}

\begin{equation}  
 {{\partial \rho\varepsilon_{\rm i}}\over \partial t}+
   {\rm div}(\rho\varepsilon_{\rm i}\bmath{v})
     = -p_{\rm i}\;\rm div \bmath{v} + \Phi - q^{\rm ie}
\end{equation}
and
\begin{equation}  
 {{\partial \rho\varepsilon_{\rm e}}\over \partial t}+
   {\rm div}(\rho\varepsilon_{\rm e}\bmath{v})
     = -p_{\rm e}\;\rm div \bmath{v} + q^{\rm ie} - q_{\rm syn}- q_{\rm br}, 
\end{equation}
where  $\bmath{v} =(v, w, v_\varphi)$ are the
three velocity components,  $\rho$ is the density, $\varepsilon_{\rm i}$ and $\varepsilon_{\rm e}$ are the specific
 internal energies of the ion and electron, $p_{\rm i}$ and $p_{\rm e}$ are the 
 gas pressures of the ion and electron,  $S=(\bf{S}_r,\bf{S}_\zeta,\bf{S}_\varphi)$ denotes
 the viscous stress tensor and $\Phi = (S\;\bigtriangledown)\bf v$ is the viscous dissipation
 rate per unit mass. The full expression of the viscous stress tensor $S$ and the dissipation rate
 $\Phi$ are given in \citet{b26}. Here, the kinematic viscosity $\nu$ is given by

 \begin{equation}
  \nu = \alpha c_{\rm s}\; \rm{min}\; [H_{\rm p},H],
 \end{equation}
  where $ c_{\rm s}, H_{\rm p}$ and $H$ are the local sound speed, the pressure  scale height and
 the disc thickness, respectively, and $\alpha$ is a dimensionless viscosity parameter which
 is usually taken to be in the range $10^{-3}\le \alpha<1$.
The pseudo-Newtonian potential  is adopted in the momentum equation. 
 $q^{\rm ei}$,  $q_{\rm br}$ and $q_{\rm syn}$  are the energy transfer rate from ions to  electrons
 by Coulomb collisions, the cooling rate of electrons by electron-ion and electron-electron bremsstrahlungs and the  synchrotron cooling rate, respectively \citep{b10,b23,b34}.
 In calculation of $q_{\rm syn}$, we give the magnetic field $B$ at each radius, assuming that
 the ratio $\beta$ of the magnetic energy density to the thermal energy density is constant
 throughout the region. 
 We neglect the radiation transport assuming that the flow is optically thin.

 In the equations (5) and (6), we assume that all of the viscous dissipation energy is given to ions
 because an ion particle is much heavier than an electron particle and  it is poorly known 
 how the viscous dissipation energy is shared between the ions and electrons. However, as a representative case, 
 we examined what result is obtained when half of the dissipation energy is shared to
 the electron and observed that the total luminosity is altered maximumly by a small factor of 1.5 due to the 
 larger electron temperature in the intermediate region on the equator, although the mass-outflow rate is not
 altered largely. This indicates that the above assumption is reasonable for our present study. 

 Recent MHD simulations show that $r-\varphi$ component of the viscous stress
 is dominated over other components. However, we used the true kinematic viscous stress which includes
 all components of the stress, in expectation of enhancement of the outflow activity through the angular
 momentum transfer and the viscous dissipation.
 From some examinations of the effects of viscous stress components,  we find that in both cases
  B and C the global features of the total luminosity and the mass-outflow rate are not altered largely
 even if only the  $r-\varphi$ component of the stress is used.

In the present two-temperature plasma with the ion temperature $T_{\rm i}$ and the electron
temperature $T_{\rm e}$, the adiabatic indices for ions ($\gamma_{\rm i}$) and electrons 
($\gamma_{\rm e}$) possess different values depending on their temperature states. In reality,
we use $\gamma_{\rm e}$ is $1.6$ for $kT_{\rm e} \le m_{\rm e} c^2$, equivalently,
$T_{\rm e} \le 5.9\times 10^9 {\rm K}(=T_{\rm c})$ where electrons are non-relativistic and
$\gamma_{\rm e}$ is 4/3 for $kT_{\rm e} \ge m_{\rm e} c^2$ where electrons are relativistic.
Here, $T_{\rm c}$ is the critical electron temperature at which electrons in the non-relativistic
state change into the relativistic state. For ions, $\gamma_{\rm i}$ is taken to be 1.6 throughout
the region because ions remain in a non-relativistic state of $kT_{\rm i} \le m_{\rm p}c^2$ \citep{b11}.

The set of partial differential equations (1)--(6) is numerically solved by a finite-difference
method under adequate initial and boundary conditions. The numerical schemes used are basically
the same as that described by \citep{b24}. The methods are based on an explicit-implicit finite
difference scheme.

\subsection{Modeling of the thick accretion flow}

We focus here on the geometrically thick and optically thin accretion discs. The low angular
momentum flow  and the advection-dominated accretion flow around the black holes belong to
this category. Using the flow parameters of the thick accretion flow, we examine the
time-dependent behaviors of the outflow which is ejected from the outer boundary. We
consider three representative cases of the accretion flow with no angular
momentum (case A), the low angular momentum flow (case B) and the advection-dominated accretion
flow (case C), respectively. In the previous work, we have studied the low angular
momentum flow model in the inviscid limit \citep{b25,b24} but we intend to re-examine here
considering the effect of viscosity. The case A is studied for initial flow with flow
parameters same as case B. For case C, we start with the self-similar solutions of the
advetion-dominated accretion flow \citep{b23,b40}.   
Following this, the self-similar solutions are given by
  \begin{equation}
  v \approx -0.367\; \alpha r^{-1/2} ,
 \end{equation}
  \begin{equation}
  \lambda \approx  0.292\; r^{1/2} ,
  \end{equation}
  \begin{equation}
  {c_s}^2 \approx  0.156\; r^{-1},
  \end{equation}
  \begin{equation}
   \rho=1.05\times 10^{-4}\; \alpha^{-1} m^{-1} \dot m r^{-3/2+s} \;\;{\rm g}\; {\rm cm}^{-3},
  \end{equation}
where $\lambda$ is the specific angular momentum, $s$ is the index parameter.
Here, $r$ is measured in units of the Schwarzschild radius $R_{\rm g}$,  $v$ and
$c_{\rm s}$ are measured in speed of light $c$ and $\lambda$ is given in units of $cR_{\rm g}$.
The mass of the black hole $m$ and the accretion rate $\dot m$ are expressed in unit of solar
mass and the Eddington rate of $1.4\times 10^{18}~{\rm gm~s^{-1}}$, respectively.
In this self-similar solution, we consider a constant mass accretion rate with the index $s=0$.
Although recent HD simulations of the thick hot accretion flow over a wide range of
$r \sim 10^4 R_{\rm g}$ \citep{b43} reveal slightly different relations compared
to the above self-similar solutions (i.e. $\rho \propto r^{-0.65}$ and $r^{-0.85}$ for $\alpha$
=0.001 and 0.01, respectively), eventually it does not affect the present study of the mass-outflow.
In Table 1, we show the flow parameters for cases A, B and C, where the flow variables at the
outer disc boundary are illustrated. The electron temperatures at the outer boundary
are approximately taken to be equal to the ion temperatures.

\begin{table*}
\centering
\caption{Flow parameters as specific angular momentum $\lambda$, adiabatic
index of ions $\gamma_{\rm i}$, mass accretion rate $\dot M\;( M_{\odot}\;{\rm yr}^{-1})$,
radial velocity $v_{\rm out}$, density $\rho$, ion temperature $T_{\rm i}$, relative disc
thickness $h/r$, viscosity parameter $\alpha$ and ratio $\beta$ of the magnetic energy density
to the thermal energy density at $R_{\rm out}=200 R_{\rm g}$ in the two-temperature model.}
\begin{tabular}{@{}cccccccccc} \hline \hline
 case&$\lambda$ & $\gamma_{\rm i}$ &$\dot M\;( M_{\odot}\;{\rm yr}^{-1})$&-$v_{\rm out}/c$  &$\rho$& $T_{\rm i}$ (K)
  & $h/r$& $\alpha$& $\beta$               \\      \hline
 ${\rm A}$    & 0.0  & 1.6 &$4.0\times 10^{-6}$
        & $5.06\times 10^{-2}$ &$5.61\times 10^{-19}$  & $1.292\times 10^9$ &0.43 & 0.0--0.1& $10^{-3}$ \\
 ${\rm B}$    & 1.35 &1.6 &$4.0\times 10^{-6}$
        & $5.06\times10^{-2}$ & $5.61\times 10^{-19}$ & $ 1.292\times 10^9$  &0.43 & $10^{-3}$--0.01& $10^{-3}$ \\
 ${\rm C}$    & 4.13 &1.6& $1.5\times 10^{-6}$
        & $2.60\times10^{-2}$ & $4.14\times 10^{-18}$ & $ 2.082\times 10^9$  & 0.50 & $10^{-3}$--0.1&$10^{-3}$ \\
\hline
\end{tabular}
\end{table*}

\subsection{Initial and Boundary Conditions}

The initial discs of  cases B (also A) and C are given by 1D solutions of the low angular momentum accretion 
 flow model \citep{b24} and the self-similar solutions (8)--(11) of the advection-dominated accretion flow, respectively. 
 Furthermore,  assuming the density distribution of $\rho \sim r^{-1}$, we construct an initial rarefied atmosphere 
 around the discs to be approximately in radially hydrostatic equilibrium.

Physical variables at the inner boundary, except for the velocities, are given by extrapolation
of the variables near the boundary. However, we impose limiting conditions that the radial
velocities are given by a fixed free-fall velocity and the angular velocities are zero.
On the rotational axis and the equatorial plane, the meridional tangential velocity $w$ is
zero, and all scalar variables must be symmetric relative to these axes. The outer boundary
at $r=r_{\rm out}$ is divided into two parts --- one is the disc boundary
through which matter is continuously entering from the outer disc with a constant accretion
rate $\dot M$ and the other is the outer boundary region above the accretion disc. We impose
free-floating conditions on this outer boundary and allow matter to eject in the
form of outflow where any inflow is prohibited. 
All flow variables at the outer boundary of the disc are kept constant always.

\section{The numerical results}

\subsection{The thick accretion flow with no angular momentum}

Case A is the simplest case of thick accretion flows. The initial flow variables of case A 
 are given by 1D solution of  the low angular momentum flow model (case B) in Table 1 but
the angular momentum is neglected here. 
 After the time integration, the upstream accreting gas falls towards
the gravitational center and soon reaches close to the inner edge of the disc. Though most
of the accreting gas is primarily swallowed into the black hole, a part of the gas is 
ejected upward along the z-axis due to the high temperature near the inner edge.
Since the ejected gas interacts with the downward accretion gas with considerable amount
of horizontal velocity component, an oblique shock is formed along the z-axis and is elongated
 in the region of  $z\sim 80 R_{\rm g}$ and $r\sim 30 R_{\rm g}$ at $t\sim 10^5 $ s.
Eventually, the upward moving gas in the shocked region is escaped from the outer boundary.
As a result, a few to several percent of the input accreting matter is ejected through the outermost
boundary.

\begin{figure}
   \begin{center}
     \includegraphics[width=80mm,height=60mm]{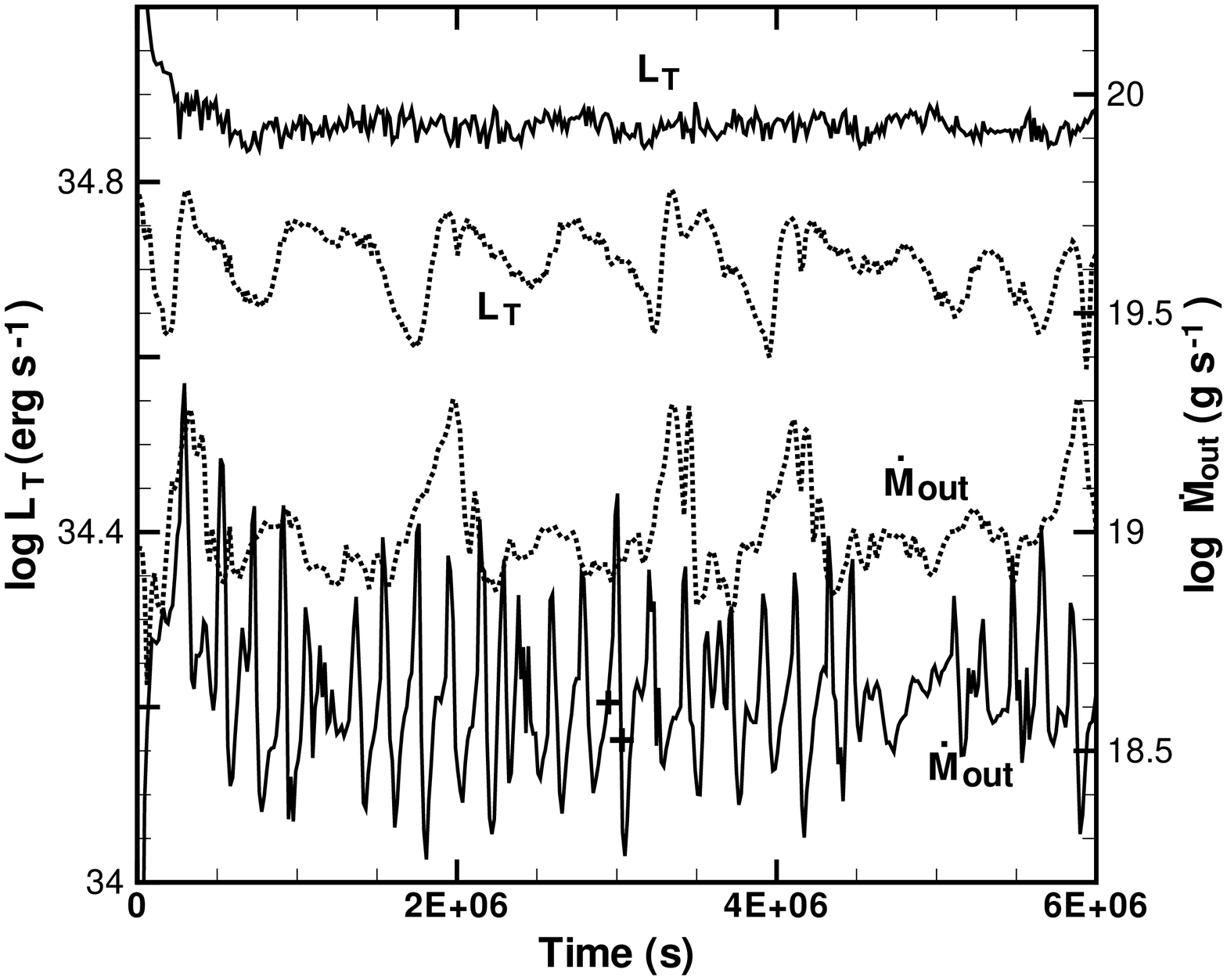}
     \caption{Time variations of total luminosity $L_{\rm T}$ (erg s$^{-1}$) and 
       mass-outflow rate $\dot M_{\rm out}$ in case A with the viscosity 
       parameter $\alpha$ = 0.0 (solid lines) and 0.1 (dotted lines).}
     \label{fig1}
 \end{center}
 \end{figure}

Fig.~1 shows the time variations of the total luminosity $L_{\rm T}$ integrated in the
computational domain $V$ and mass-outflow rate $\dot M_{\rm out}$ ejected from the atmospheric
outer boundary $S$  for case A.
Here, $L_{\rm T}$ and $\dot M_{\rm out}$ are given by

 \begin{equation}
     L_{\rm T} =\int (q_{\rm br} + q_{\rm syn}) {\rm d} V,
 \end{equation}
 
\noindent and

 \begin{equation}
    \dot M_{\rm out} = \int {\rho}_{\rm out}\bmath{v_{\rm out}}{\rm d} \bmath{S},
 \end{equation}
where the suffix `out' shows the value at the atmospheric outer boundary.
The solid and
dotted lines represent the cases with $\alpha = 0.0$ and $0.1$, respectively.
We find that the luminosities and the mass-outflow rates vary quite irregularly.
In particular, the variation for first one is less than 20 $\%$ and for second one is by
a factor of few. Here, the luminosities are mainly dominated by the bremsstrahlung emission.
The mass-outflow rates increase with the increasing viscosity and conversely the luminosities
decrease especially in the case of the maximum viscosity with $\alpha=0.1$.
Due to the viscous heating, ion temperature becomes higher and therefore, the shock wave near the
z-axis is pushed outward. Hence, the outflow over the oblique shock is enhanced as seen in Fig.~2.
On the other hand, the electron temperature behind the shock becomes lower when the shock is
pushed away. As the dominant energy emission process is due to the free-free emission which mostly
depends on the electron temperature behind the shock, the luminosity evetually becomes lower by 
a small factor.

 We wonder why the mass-outflow rates are considerably modulated chaotically in spite of the
small modulations of the luminosity. To understand the outflow activity,
it is useful to follow the animation of the time-dependent density and temperature contours.
Fig.~2 shows the snapshots of the animation over a typical cycle in the time variation of
the mass-outflow rate, where the temperature contours at $t/10^6= 2.899$,  2.929,  2.960,  2.990, 
3.005 and 3.036 s are shown.  
The large mass-outflow activity begins at the elongated region of the shocked surface.
Firstly, a hot and rarefied blob with small size is formed at the upper edge ($z\sim 80$ and 
$r\sim 40$) of the shocked region. The hot blob grows up in the outward direction
and starts disintegrating into several small blobs. Ultimately, these blobs further continue to
expand outward. The maximum mass-outflow rate is attained at the time of Fig.~2(d).
Thus, the unstable mass-outflow is observed even in the case of the
thick accretion flow with no angular momentum.

\begin{figure*}
\begin{center}
     \includegraphics[width=12cm,height=8cm]{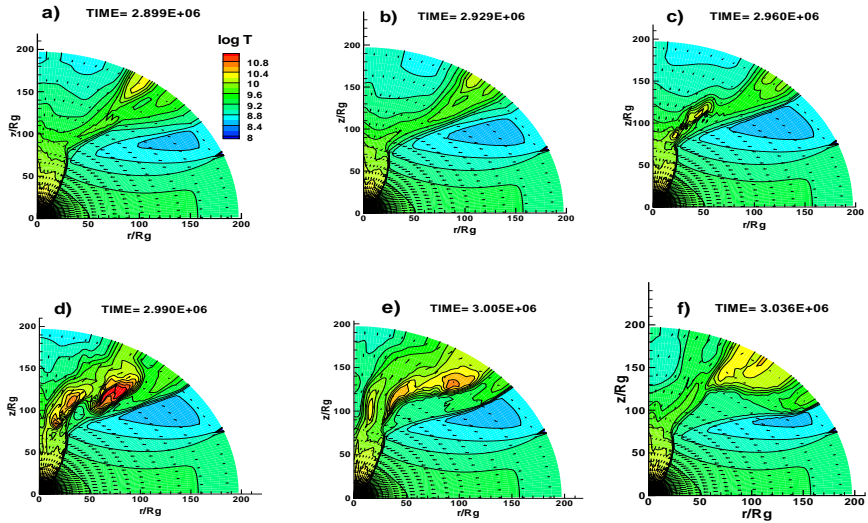}
     \caption{Subsequent snapshots of temperature contours at phases of $t/10^6$= 2.899,  2.929,  2.960,
  2.990, 3.005 and 3.036 (s) during an eruptive mass ejection in the curve of $\dot M_{\rm out}$ with $\alpha=0.0$ 
  of Fig.~1. The phases of the snapshots (a) and (f) are shown by two crosses in the solid line $\dot M_{\rm out}$
  of Fig.~1 }  
     \label{fig2}
 \end{center}
 \end{figure*}

\subsection{The low angular momentum accretion flow}

The low angular momentum accretion flows without viscosity were examined
in the previous works \citep{b25,b24}.  Using same sets of  the model parameters of the specific
angular momentum $\lambda$ and the specific total energy $\epsilon$ which were estimated from 
the analysis of stellar wind of nearby stars around Sgr A* \citep{b8,b19}, we showed that 
the irregularly oscillating shocks are formed in the inner region and consequently the
luminosity and mass-outflow rate are modulated within a factor of few and several, respectively. 
In this work, we study the low angular momentum flow model including viscosity and observe that
the characteristics of the low angular momentum flow are not altered largely compared to the inviscid
 flow, when viscosity parameter $\alpha$ is small as $\sim 10^{-3}$. 
Similarly to case A,
an oblique shock surface is formed due to the interactions between the upward moving gas and
the downward accreting gas. Since the flow
is rotating around the z-axis, the shock surface on the equatorial plane is pushed away
in the outward direction due to the centrifugal force and is oscillated irregularly at
$r/R_{\rm g} \sim 20$ -- 50. During shock oscillations, a small hot blob is occurred in the shocked
region near the equatorial plane, grows upward, undergoes to become maximal at
the upper edge of the shocked region and finally attains the outer boundary. 
The evaluation of this hot blob seems to be suitable to explain the episodic mass-outflow activity. 

Fig.~3 represents the time variations of the total luminosity ${L_{\rm T}}$ erg s$^{-1}$,
the mass-outflow
rate $\dot M_{\rm out}$, the mass-inflow rate $\dot M_{\rm in}$ at the inner
edge and the shock position $R_{\rm s}/R_{\rm g}$ on the equatorial plane for
case B with $\alpha=10^{-3}$. The mass-outflow rate attains 10 $\%$ of the input accretion
rate and varies by a factor of 2 -- 3 whereas the luminosity is modulated
by a factor of 1.5. The shock position on the equatorial plane oscillates irregularly around
$r\sim 35R_{\rm g}$.
The large mass-outflow activity occurs roughly in
regular intervals of the order of few days.
Fig.~4 shows the contours of ion temperature $T_{\rm i}$ at
$t= 1.83\times 10^6$ s for case B with  $\alpha=10^{-3}$. The time is indicated by `+' on
the curve  $\dot M_{\rm out}$ in Fig.~3. The shock surface is illustrated
by the black thick lines which are located at $r\sim 35~R_{\rm g}$ on the equatorial plane
and is elongated obliquely. The rarefied hot funnel region is shown by the red zone
along the z-axis. 

On the other hand, when $\alpha$ is as large as 0.1, the shock wave formed initially in the inner region
that expands outward and finally disappears from the outer boundary. At the later phases, a large scale
convective motion is prevailed in the entire accretion flow. As a result, most of the input accreting
matter is ejected from the outer boundary and the mass-outflow rate is modulated by a factor of few.

\begin{figure}
     \includegraphics[width=86mm,height=66mm]{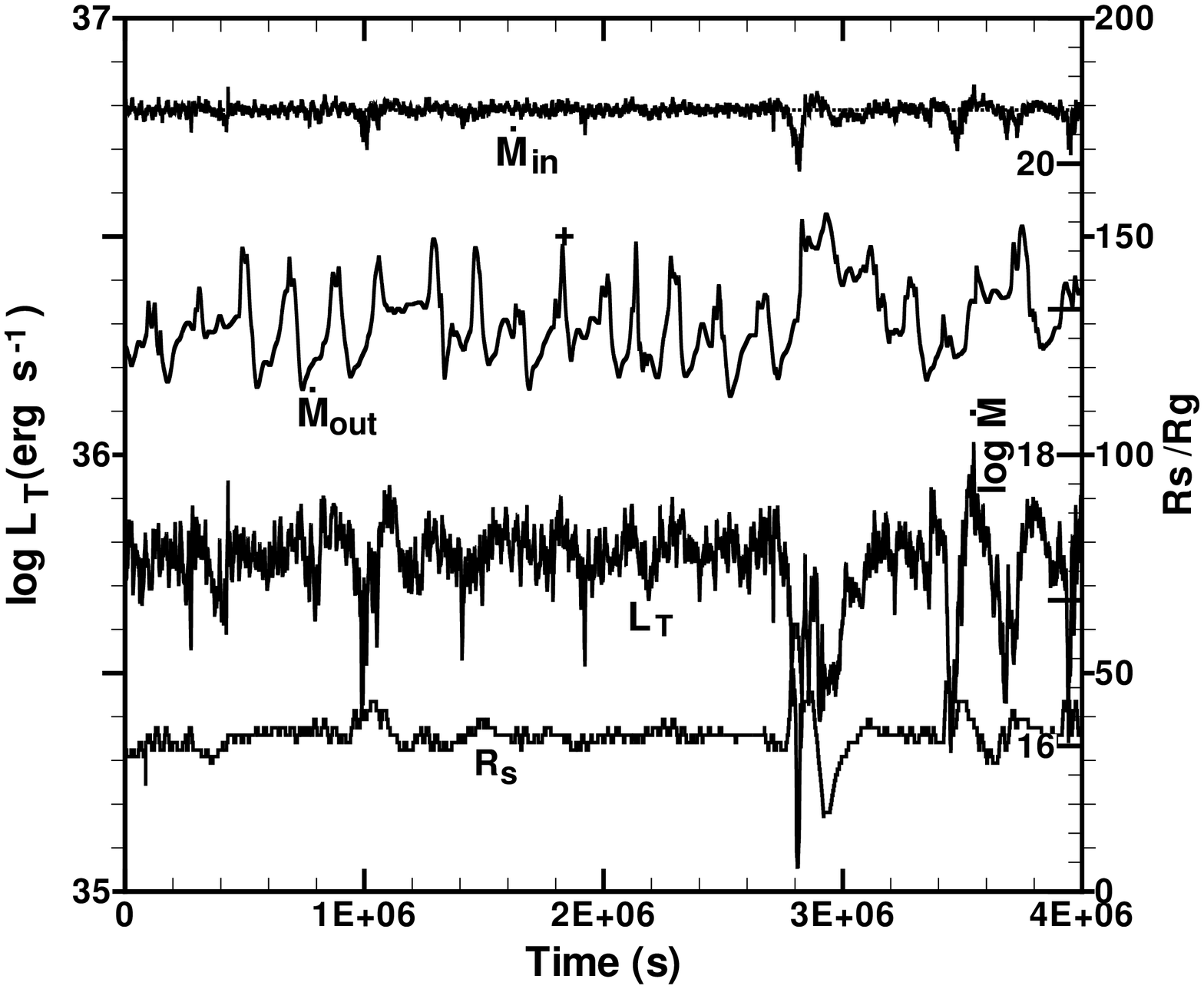}
     \caption{Time variations of total luminosity $L_{\rm T}$ (erg s$^{-1}$), 
    mass-outflow rate $\dot M_{\rm out}$ (gm s$^{-1}$),
       mass-inflow rate $\dot M_{\rm in}$ at the inner edge and shock position
       $R_{\rm s}/R_{\rm g}$ on the equatorial plane for case B with
       $\alpha=10^{-3}$.}
     \label{fig3}
 \end{figure}

\begin{figure}
     \includegraphics[width=86mm,height=66mm]{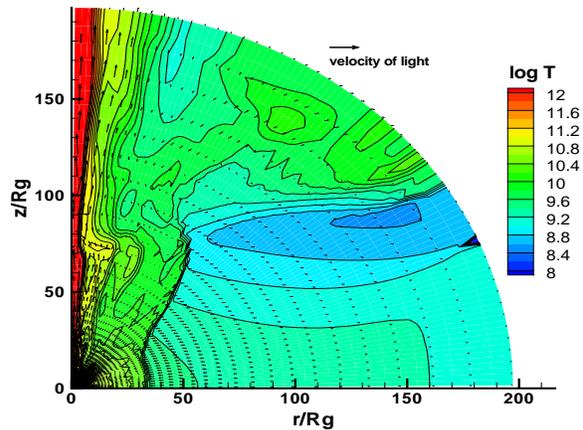}
     \caption{Contours of ion temperature $T_{\rm i}$ with velocity vectors at
     $t=1.83\times 10^6$ s of case B with $\alpha=10^{-3}$. The evolutionary phase
     is denoted by `+' in the curve $\dot M_{\rm out}$ of Fig.~3.
     The velocity of light is indicated by the upper arrow.}
     \label{fig4}
 \end{figure}

\subsection{The advection-dominated accretion flow}

Although the initial flow variables of case C are obtained by the 
self-similar solution of the advection-dominated flow, the evolution of the flow 
at the initial state is similar to that of case B. The upstream
accreting gas falls towards the gravitational center and the outflows
are generated from the surrounding region of the inner edge. These
outflowing gas interacts with the downward accreting gas, resulting in the formation of
oblique shock. However, since the flow has larger angular momentum than that
in case B, the shock on the equatorial plane continues to
move outward gradually and finally reaches the outer disc boundary. As a result,
the phenomenon of shock formation is never recovered. After a long integration time
 more than $\sim 100$ orbital period at the outer boundary, the entire flow seems
to settle down in a steady state, although the luminosity is still modulated by a
very small factor. On the other hand, the mass-outflow rate remains as large as
$10\%$ of the input accretion rate and varies by a factor of ten or so.

Fig.~5 shows the time variations of the total luminosity ${L_{\rm T}}$ erg s$^{-1}$ and
the mass-outflow rate $\dot M_{\rm out}$ g s$^{-1}$ for case C with $\alpha=0.1$.
In the plot, dashed horizontal line denotes the input accretion rate $\sim 10^{20}$
g s$^{-1}$. It should be noticed here that the mass-outflow ejected from the outer boundary
surface  with the polar angle $\zeta \geq 60^{\circ}$ is presented in Fig.~5
due to the following reason. We imposed free-floating conditions on the outer 
atmospheric boundary which allow only for outflowing matter. However, recent
HD simulations of the hot accretion flow on large scales, namely,
from the black hole horizon to  10 Bondi radius \citep{b15} indicate that the
above boundary conditions may be invalid. This is because convective motions
prevail even in the outer boundary region used in our simulations and accordingly
there may exist outer flows with not only positive velocities but also negative
ones at the outer boundary. On contrary, the boundary conditions adopted in this
work seems to be justified in the polar region with $\zeta \geq 60^{\circ}$, where the
high velocity jets with only positive radial velocities appear in this conical region.

  \begin{figure}
     \includegraphics[width=86mm,height=66mm]{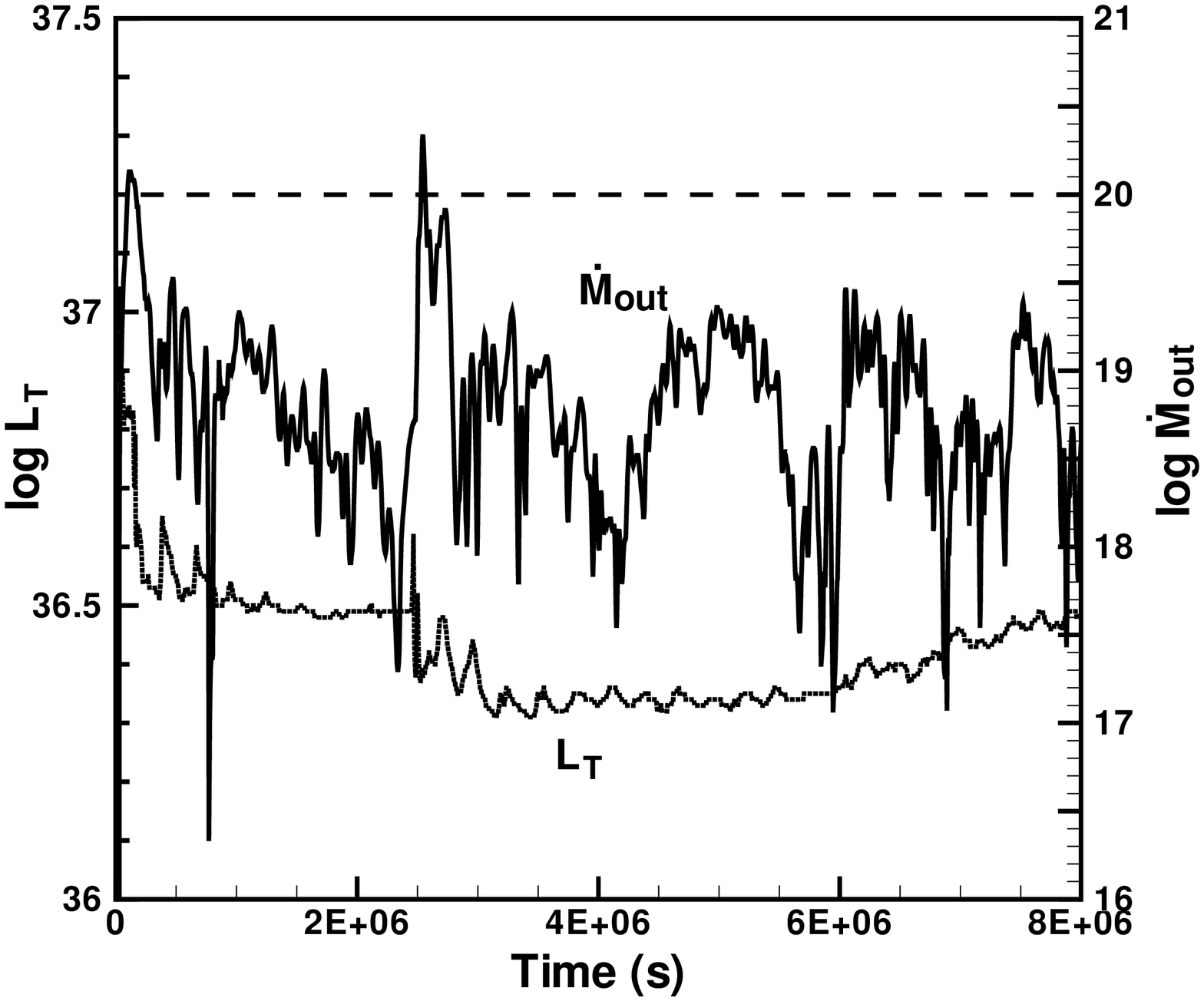}
     \caption{Time variations of total luminosity $L_{\rm T}$ (erg s$^{-1}$) and 
     mass-outflow rate $\dot M_{\rm out}$ in case C with $\alpha=0.1$, where the
     dashed horizontal line denotes the input accretion rate.}
     \label{fig5}
  \end{figure}

In Fig.~6, we present the contours of the density at
$t = 5.0\times 10^6$ s for case C with $\alpha=0.1$. The corresponding 
contours of ion temperature  are depicted in Fig.~7, where the velocity
vectors are given by unit vectors. Here we find that the flow is unstable
against convection and many convective cells are found in the inner region.
It is remarkable that a torus disc with an inner edge of $r \sim 3R_{\rm g}$
and a concentric center at $r\sim 6R_{\rm g}$ is formed in the advection-dominated
accretion flow.

\begin{figure}
     \includegraphics[width=80mm,height=60mm]{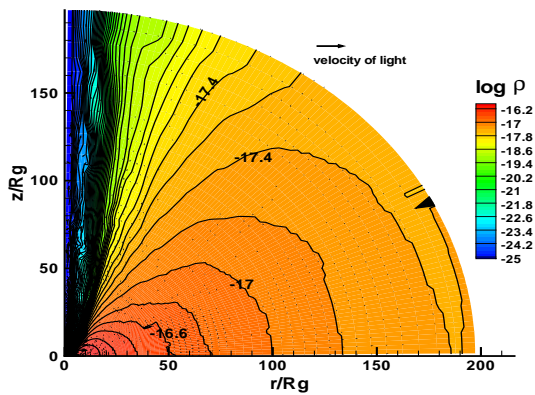}
     \caption{Contours of density  $\rho ({\rm g}$ ${\rm cm}^{-3})$ and 
     velocity vectors at $t = 5.0\times 10^6 \;$ (s) for case C. 
     The velocity of light is indicated by the upper arrow.}  
     \label{fig6}
\end{figure}
 
  \begin{figure}
     \includegraphics[width=80mm,height=60mm]{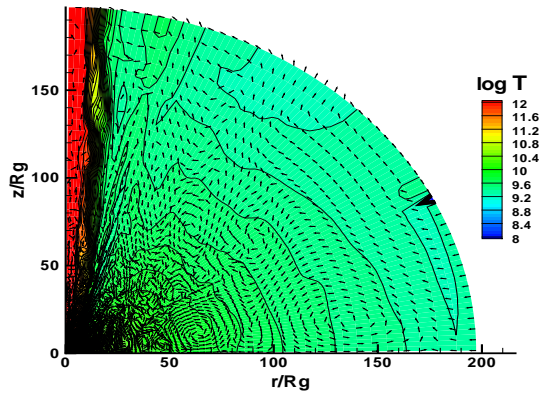}
     \caption{Same as Fig.~6 but for the ion temperature $T_{\rm i}$ with unit
     velocity vectors, where many convective cells are found in the inner region.  }  
     \label{fig7}
  \end{figure}

In Fig.~8, we show the variations of density $\rho$, ion temperature
$T_{\rm i}$, electron temperature $T_{\rm e}$ and radial velocity $v$ on the equatorial
plane as function of radial coordinate. Note that the density and the temperatures on
the equator have their maximum values at $r\sim 6R_{\rm g}$ and decrease sharply inward.
At $r \sim 3R_{\rm g}$, the temperatures start increasing discontinuously although density
continues to decrease even further. In Fig.~9, we zoom the inner edge of the torus disc
to represent the contours of density with velocity vectors at the same phase as Figs.~6 and~7
for clarity. It is clear from the figures that the torus disc with the  center
at $r\sim 6R_{\rm g}$ has the inner edge of the disc at $r \sim 3R_{\rm g}$. We observe that the outflow
begins at $r \sim 3R_{\rm g}$ on the equator and goes up along the outer surface of
the torus disc.  In the region $r < 3R_{\rm g}$, the gas falls toward the black hole and the
density at the torus edge becomes very small and  the flow has a discontinuous
surface of the temperature there.

In order to understand the irregular oscillation of mass-outflow, we examine
the time-dependent animations of density and temperature contours with velocity vectors. 
Then, we notice the formation of a series of hot blobs near the inner edge of the torus and their
subsequent evolution along the outer surface of the torus. These hot blobs are originated
in the innermost region between the inner edge ($r/R_{\rm g} \sim 3$) of the torus disc and
the inner edge ($r/R_{\rm g}=1.5$) of the flow.

 \begin{figure}
  \includegraphics[width=80mm,height=60mm]{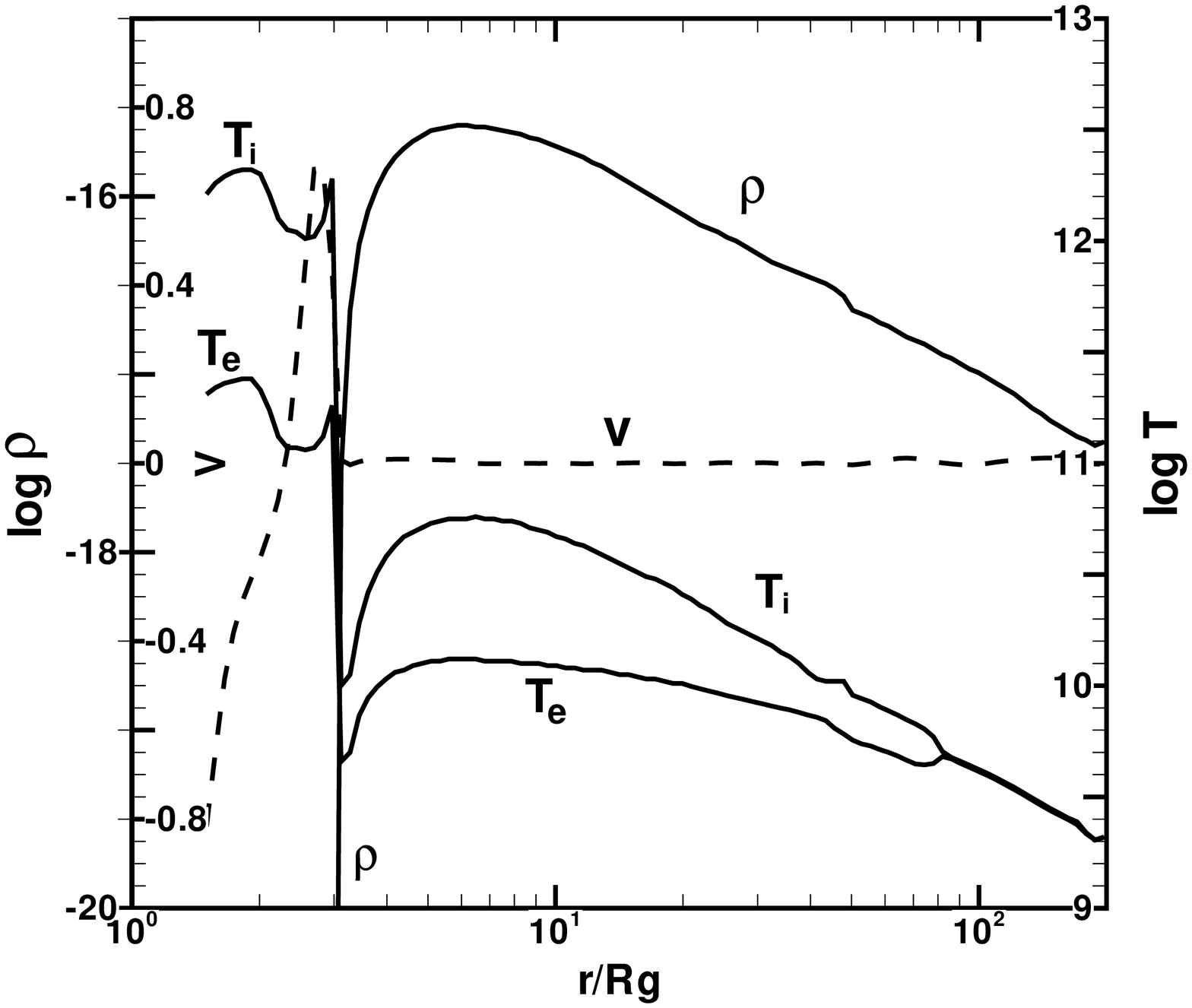}
    \caption {Variations of density $\rho$ (${\rm  g}\; {\rm cm}^{-3}$),
    ion temperature $T_{\rm i}$, electron temperature $T_{\rm e}$ and radial velocity $v$
    on the equatorial plane as observed in case C.}
     \label{fig8}
  \end{figure}
 
 \begin{figure}
     \includegraphics[width=80mm,height=60mm]{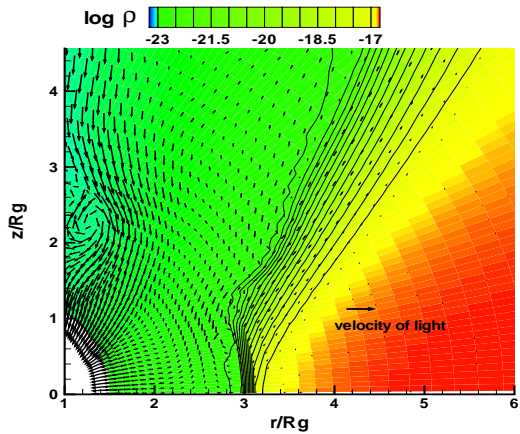} 
     \caption{Same as Fig.~6, but near the inner edge, where at $r\sim 3R_{\rm g}$
     the mass-outflow begins upward along the disc surface. }
     \label{fig9}
  \end{figure}

We examine another case with $\alpha=10^{-3}$ for the advection-dominated accretion flow.
In this case, we again find the similar results as seen for $\alpha=0.1$,
i.e., the large modulation of the mass-outflow rate, the small modulation of the
luminosity and the torus disc formation. The mass-outflow rate varies irregularly by more
than a factor of ten and roughly occurs in every several hours. A series of
hot blobs is also generated in the innermost region
 and grows up along the boundary zone between the outflow and the outer
surface of the disc. The torus disc has the  center at $r\sim 10R_{\rm g}$ and
the inner edge of the disc  at $r \sim 6R_{\rm g}$. The total
luminosity is obtained as $\sim 10^{37}$ erg s$^{-1}$ which is one order of magnitude
larger than the case with $\alpha=0.1$. This is due to the fact that the input
density at the outer boundary is much larger as seen from equation (11).

\section{Kelvin-Helmholtz instability as the origin of the unstable nature of the outflows}

We have examined that the mass-outflows in the geometrical thick accretion flows 
such as cases A, B and C  are generally unstable and the mass-outflow
rates vary by more than a factor of few  up to ten. In cases of A
and B, the apparent origin of the outflow activity is ascribed to the
formation of hot blobs and its evolution toward the outer boundary. We have examined such
modulations of the mass-outflow rate in other simulations of 2D rotating accretion flows 
around a stellar-mass ($10 M_{\odot}$) and a supermassive  ($10^6 M_{\odot}$) black holes
over a wide range of input accretion rates as $10^{-7} \le \dot M/\dot M_{\rm E} \le 10^{-4}$
\citep{b28}. In addition, examining 2D simulations for the disc of SS 433
at super-Eddington luminosities, we found remarkable modulations of the accretion rate
near the inner edge that results in the formation of recurrent hot blobs with high temperatures
and low densities at the disc plane. Based on this observation, we proposed that this may
explain the massive jet ejection as well as the QPOs phenomena observed in SS 433 \citep{b27}.
 
What is the intrinsic origin of the unstable nature of the outflows which is relevant to the hot
blobs?  In this respect, we consider shear instability or
Kelvin-Helmholtz instability that may lead to the hot blobs and the winding waves of density.
The shear instability and the Kelvin-Helmholtz instability can generally occur when there
exists a velocity shear in a single continuous fluid or if there is a velocity difference
across the interface between two fluids. Finally, the theory predicts the onset of instability
and transition to the turbulent flow in fluids of different densities moving at various speeds
\citep{b7}. Certainly, in cases A and B, there exists an interface between the outgoing flow
and the downward accreting gas as the shock surface with discontinuous
velocities and densities and the hot blobs are formed along this interface and are evolved toward
the upper shocked region. For case C, the hot blobs exist along the interface between the outflow
and the outer surface of the torus disc. Therefore, the Kelvin-Helmholtz instability generated
at the interface may be developed into the hot blobs and the winding waves.

For the sake of simplicity, let us consider an extreme type of stratification, namely a plane
parallel two-layer system in which a layer of fluid with velocity $v_1$ and density ${\rho}_1$
floats over another layer with velocity $v_2$ and density ${\rho}_2$ where a gravitational force
$g$ acts vertically to the layers. Then, gravity waves propagate through the
interface separating the two layers and grow in time. This leads to overturning in the 
vicinity of the interface and mixing over a height $\delta H$ along the z-axis vertical to the 
layers, which is called the Kelvin-Helmholtz instability. 
In this simple picture, the growth of the Kelvin-Helmholtz
instability is determined by a criterion of Richardson number \citep{b7,b31},
 
\begin{equation}
  {\rm R}_{\rm i} \equiv { {-g (d\rho/dz)} \over {{\rho(dv/dz)^2}}} < {1 \over 4}.
\end{equation}

\noindent Then, we have the Richardson number approximately as

\begin{equation}
{\rm R}_{\rm i} \simeq  { {-g \delta H \mid\rho_2-\rho_1\mid } \over {\bar{\rho}(v_2-v_1)^2}}.
\end{equation}

In Fig.~2, we find the oblique shock front present in the accretion flow.
 Regarding the shock front and $\delta H$ as the interface mentioned above and the shock
 thickness, respectively, we estimate the Richardson number. 
Across the interface, if the velocity difference parallel to the shock
surface is not large enough, the criterion described in Eq. (14) is never satisfied.
Fig.~10 illustrates the shock surface (thick line) as in Fig.~2 and the geometry of the
shock front with thickness $\delta H$ where ${\rm S}(r,\zeta)$ and
${\rm T}(r,\zeta+\delta \zeta)$ are the intersection points of the shock front with a
circle of radius $r$ centered at the origin, $\omega$ is the angle between the gravitational
force direction and the tangent to the shock surface at point $\rm S$, $\eta$ is the angle
between the tangent at point $\rm S$ and x-axis. With this, the Richardson number in units of
the Schwarzschild radius and the speed of light is given by,

\begin{equation}
R_{\rm i} \simeq  {1\over 2} { {\mid{\rho_2-\rho_1}\mid } \over \bar{\rho}} {\delta \zeta \over r}{{\rm sin~} \omega~{\rm cos~} (\omega -{\delta \zeta \over 2})
  \over (v_2-v_1)^2 },
\end{equation}
where  $v_1$ and $v_2$ are the components of pre-shock and post-shock velocities
parallel to the tangent at $\rm S$. The physical variables at the points $\rm S$ and $\rm T$
are determined approximately from the numerical data of the flow.
As an example, we take a point on the shock front in Fig.~2a., whose radius is 67.4$R_{\rm g}$.
For this point,  in Fig.~11, we
show the variation of density $\rho$, radial velocity $v_{\rm r}$ and azimuthal velocity
$v_{\rm \varphi}$ as function of polar angle $\zeta$ measured from the equatorial plane. 
The shock transition is found here in the region of $63.9^\circ \leq \zeta \leq 67.5^\circ$ and 
 $\delta \zeta \sim 3.6^\circ$. From the numerical data, furthermore,
 we know  $\rho_1$,  $\rho_2$, $(v_{\rm r})_1$,  $(v_{\rm r})_2$,
$(v_{\rm \varphi})_1$, $(v_{\rm \varphi})_2$, $\omega(=\eta-\zeta)$,
$v_1=(v_{\rm r})_1 {\rm cos~} \omega +(v_{\rm \zeta})_1 {\rm sin~} \omega$, 
$v_2=( v_{\rm r})_2 {\rm cos~} (\omega-\delta\zeta) +(v_{\rm \zeta})_2 {\rm sin~} (\omega-\delta\zeta)$  
and  finally estimate $R_{\rm i}$. For case C, we do not find the shock interface found in cases A and B.
 However, there exists a discontinuous boundary zone between the torus disc and the outflowing gas
 where change of temperature and velocities occur very sharply as observed in Figs.~8 and 9.
 We apply the above method at this boundary zone to obtain the Richardson number.
  
In Fig.~12, we plot the distribution of the Richardson number
$R_{\rm i}$ as a function of radial distance $r/R_{\rm g}$ obtained for the
 oblique shocks in Fig.~2(a) of case A (open squares) and in Fig.~4 of case B (diamonds), and for the boundary zone around the torus disc as in Fig.~9 of case C (closed circles). This shows that
the criterion (Eq. 14) for Kelvin-Helmholtz instability is not satisfied
near the equatorial plane but is established apart from the equator for all
cases. This suggests that the unstable property of the outflows would possibly
be ascribed to the Kelvin-Helmholtz instability.
 
\begin{figure}
\begin{center}
     \includegraphics[width=80mm,height=60mm]{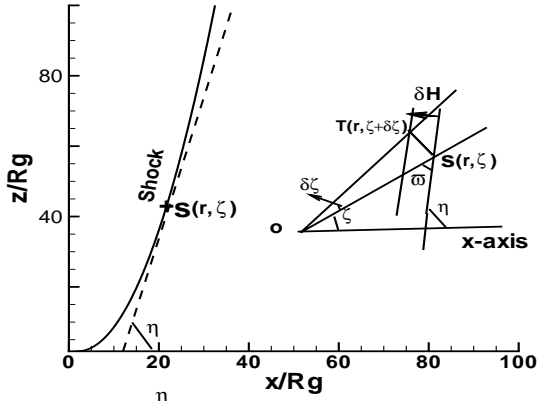}
    \caption{Geometry of shock surface with shock front thickness $\delta H$ found in
    Fig.~2, where ${\rm S}(r,\zeta)$ and ${\rm T}(r,\zeta+\delta \zeta)$ are the
    intersection points of the shock front with a circle of radius $r$ centered at
    the origin, $\omega$ is the angle between the gravitational force
    direction and the tangent to the shock surface at point $\rm S$, $\eta$ is the
    angle between the tangent at the point $\rm S$ and x-axis.}
  \label{fig10}
 \end{center}
 \end{figure}

\begin{figure}
\begin{center}
     \includegraphics[width=80mm,height=60mm]{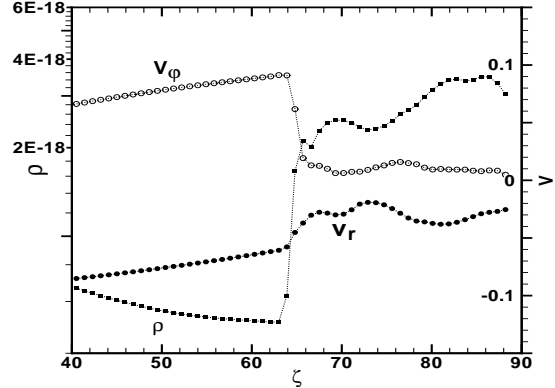}
    \caption {Plot of density $\rho$ ({\rm g} {\rm cm}$^{-3}$), radial
    velocity $v_{\rm r}$ and azimuthal velocity $v_{\rm \varphi}$ as function of polar
    angle $\zeta$ at $r/R_{\rm g}=67.4$ on the shock front in Fig ~2a, where the shock transition is found in the
    region of $63.9^\circ \leq \zeta \leq 67.5^\circ$ and $\delta \zeta \sim 3.6^\circ$.}
  \label{fig11}
 \end{center}
 \end{figure}

\begin{figure}
\begin{center}
     \includegraphics[width=80mm,height=60mm]{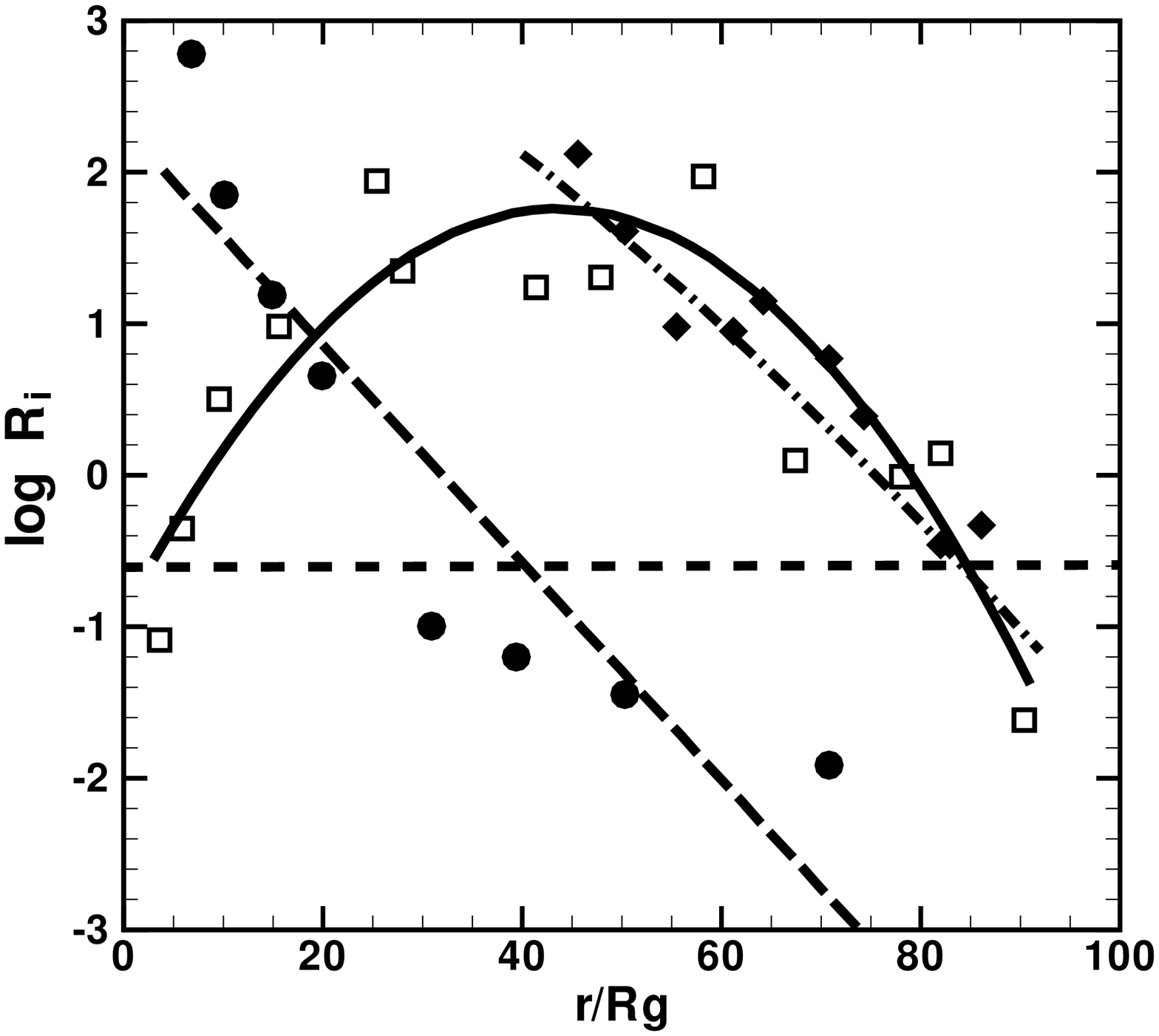}
    \caption {Variation of the Richardson number $R_{\rm i}$ as function of
    radius on the shock surface in Fig.~2(a) of case A (open squares), in Fig.~4 of case B
    (diamonds) and on the boundary zone around the torus disc in Fig.~9 of case C (circles).
    The corresponding fitted curves are shown by thick solid, dash-dotted and long-dashed
    lines, respectively. The horizontal dashed line denotes the critical value 
    $R^{cr}_{\rm i}=0.25$ for the Kelvin-Helmholtz instability.}
  \label{fig12}
 \end{center}
 \end{figure}

\section{A possible model to the flares of  Sgr A*}

The supermassive black hole candidate Sgr A* possesses a quiescent state and a flare state. 
The flares of Sgr A* have been detected in multiple wavebands from radio, sub-millimetre, IR
to X-ray. The unstable mass-outflow activities examined in this work 
demonstrate both small and large amplitude modulations of the total luminosity and the mass-outflow
rate which seems to be relevant for explaining the flaring phenomena observed in  Sgr A*.
The total luminosity found in cases B and C satisfactorily explains the
existence of low luminosity of $\sim 10^{36}$ erg s$^{-1}$ for Sgr A*. Similarly, the
irregularly eruptive mass-outflows may correspond to the flare activities of Sgr A*.

The observations of Sgr A* show that the duration of the X-ray and IR flares are typically
of 1-3 hours and the flare events usually occurs few times per day. 
In the simulation of cases B and C, we find that the duration and the interval
between the successive eruptive mass-outflows roughly are in agreement with the flares of
Sgr A*. But, the small modulations of luminosity does not correspond to
the large amplitudes of the flare intensities observed at various wavelengths, because the
amplitudes at radio, IR and X-ray are roughly varied by factors of 1/2, 1-5 and 45,
respectively \citep{b12,b13,b9,b16,b17,b35,b44,b45}. It is well known
that the sub-milimeter flare lags the X-ray flare roughly by two hours. From these facts,
we propose a possible model of the unstable mass-outflow applicable to the flare phenomena
of Sgr A*. This is basically a scenario of the temporal sequence of disc-jet
coupling observed in the X-rays, IRs and radio wavelengths of the microquasar GRS 1915+105
\citep {b18}. Here, we consider that the total emission in our models corresponds to the permanent
emission of the quiescent state of Sgr A*, whereas only the emission of Sgr A* flare
is produced at the distant region far away from the Sgr A* through the interaction between the
surrounding dense interstellar matter and the high-velocity jets. These jets are originated in
the unstable mass-outflows as examined here. 
The ejected wind carries sufficient outflowing mass of rate
$\sim 10^{19}$ -- $10^{20}$ g s$^{-1}$
and a total kinematic energy of $\sim 10^{36}$ --$10^{37}$ erg s$^{-1}$ to reproduce the flare
phenomena. However, one needs to incorporated other mechanisms also so that the
high-velocity wind could be accelerated up to the relativistic speed. This
is because the outflow velocities at the outer boundary in cases B and C are
limited to sub-relativistic regime as $\sim 0.01-0.1 c$.  
 
 When we consider  the flare activities of Sgr A*,  it is required for us to treat MHD model because it is widely believed  that the existence of the magnetic field is inevitable
 from the spectral analyses of Sgr A* and that magnetic turbulence induced by magneto-rotationa
l instability is responsible for the angular momentum transfer.
 In this respect, by analogy with the coronal mass ejection on the Sun, Youan et al. (2009) proposed
 a MHD model for the formation of episodic jets from black holes associated 
with the closed magnetic fields in an accretion flow. 
 In their model, shear and turbulence of the accretion flow deform the field and result in th formation
 of a flux rope in the corona and the flux rope is ejected through the magnetic reconnection process 
 in a current sheet under the magnetic field.

Although the unstable outflows examined here are restricted to the non-magnetized accretion flow,
 we notice that the comparative studies of 2D hydrodynamical and magneto-hydrodynamical 
accretion flows from a same thick torus around black holes  show  many similarities such as
 highly time-dependent accretion rate and Kelvin-Helmholtz instability
in spite of  the different mechanism
 of angular momentum transfer \citep{b33,b32}.
 They  also show that, for strong magnetic field, a small fraction of outflows near the pole is escaped 
 as  a powerful MHD wind, whereas the outflows in the hydrodynamical simulations are bound.
 This suggests that the unstable nature of the outflows from the thick accretion flow 
 may be  furthermore enhanced in 2D magneto-hydrodynamical simulation.

\section{Summary and discussion}

We examined the mass-outflow in the hot geometrically thick accretion flows 
such as the low angular momentum accretion flow and the advection-dominated accretion flow
around black holes, using 2D time-dependent HD calculations. The results are
summarized as follows.

(1) In the low angular momentum flows, the inward accreting matter on the equatorial plane
interacts with the outward moving gas and the centrifugally supported oblique shock is formed
along the interface of both the flows provide the viscosity parameter $\alpha$
is as small as $\alpha \le 10^{-3}$. The hot blobs, which lead to the eruptive mass-outflow,
are generated at the inner shocked region and grow up toward the outer boundary. 

(2) The advection-dominated accretion flows attain finally to a torus disc with the inner edge of
 the disc at $3 \le r/R_{\rm g} \le 6$ and the center at $6R_{\rm g} \le r \le 10R_{\rm g}$. 
 The hot blobs are formed near the inner edge of the torus disc and grow up along the outer surface of the disc. 
 
(3) In both models, the mass-outflow is unstable and eruptive in nature. As a result,
the luminosity and the mass-outflow rate are modulated irregularly and especially the relative
modulation amplitudes $\delta \dot M_{\rm out}/\dot M_{\rm out}$ of the mass-outflow rate are
as large as a few to ten.
 
(4) The origin of the unstable mass-outflow is ascribed to the Kelvin-Helmholtz instability
which is ascertained by examining the Richardson number criterion (Eq. 14) 
for Kelvin-Helmholtz instability at the inner region of the flow.

(5) The flare phenomena of Sgr A* may be explained by the unstable mass-outflow examined here, 
together with the scenario of the temporal sequence of disc-jet coupling observed in the
X-rays, IRs and radio wavelengths of the microquasar GRS 1915+105 \citep{b18}.
     
The unstable mass-outflow seems to be inevitable as far as we consider the thick accretion
flows with small angular momentum around the black holes. The present study shows that both
the models of the low angular momentum accretion flow and the advection-dominated accretion
flow produce the eruptive mass-outflows. These models would reproduce the
reasonable spectra fitted to the observed ones of Sgr A* if we take in the appropriate model
parameters, such as the accretion rate $\dot M$, the viscous parameter $\alpha$,
the magnetic field strength $\beta$ and a hybrid electron population of thermal and nonthermal
particles, following the idea and method described by \citet{b41,b42}.
Further, the evidence of angular momentum would confirm the validity of the
models. Moreover, it is required to perform magneto-hydrodynamical research
instead of only the hydrodynamical one. This is because the magnetic field will play important
roles not only in the small-scale region of the accretion flow but also the large-scale of
the interstellar space.

\label{lastpage}


\begin{thebibliography}{99}
\bibitem[\protect\citeauthoryear{ Abramowicz et al.}{1988}]{b1}
  Abramowicz M. A.,  Czerny, B., Lasota, J. P., Szuszkiewicz, E.,  
  1988, ApJ, 332, 646 

\bibitem[\protect\citeauthoryear{Abramowicz et al.}{2002}]{b2}
  Abramowicz M. A.,  Igumenshchev, I. V., Quataert, E., Narayan, R., 2002, ApJ., 565, 1101 


\bibitem[\protect\citeauthoryear{ Begelman}{2012}]{b3}
  Begelman, M. C., 2012, MNRAS, 420, 2912


\bibitem[\protect\citeauthoryear{Blandford \& Begelman}{1999}]{b4}
  Blandford, R. D., Begelman, M., C., 1999, MNRAS, 303, L1

\bibitem[\protect\citeauthoryear{Blandford \& Begelman}{2004}]{b5}
  Blandford, R. D., Begelman, M., C., 2004, MNRAS, 349, 68


\bibitem[\protect\citeauthoryear{Bu et al.}{2013}]{b6}
 Bu, D., Yuan, F., Wu, M., Cuadra, J., 2013, MNRAS, 434, 1692

\bibitem[\protect\citeauthoryear{Chandrasekhar}{1961}]{b7}
  Chandrasekhar S., 1961, Hydrodynamic and Hydromagnetic Stability,
 Oxford University Press, Oxford


\bibitem[\protect\citeauthoryear{Czerny \& Mo\'{s}cibrodzka}{2008}]{b8}
Czerny B., Mo\'{s}cibrodzka M., 2008, J. Phys. Conf. Ser.,131, 012001

\bibitem[\protect\citeauthoryear{Eckart et al.}{2006}]{b9}
Eckart A., Sch\"{o}del R., Meyer L., Trippe S., Ott T., Genzel R., 
2006, A\&A, 455, 1




\bibitem[\protect\citeauthoryear{Esin et al.}{1996}]{b10}
 Esin A. A., Narayan R., Ostriker E., Yi I., 1996, ApJ, 465, 312

 
 
\bibitem[\protect\citeauthoryear{Fukue}{1986}]{b11}
 Fukue, J., 1986, PASJ, 38, 167


\bibitem[\protect\citeauthoryear{Genzel et al.}{2003}]{b12}
Genzel R., Sch\"{o}del R., Ott T., Eckart A., Alexander T., 
Lacombe F., Rouan D., Aschenbach B., 2003, Nat, 425, 934


\bibitem[\protect\citeauthoryear{Ghez et al.}{2004}]{b13}
Ghez A. M. et al., 2004, ApJ., 601, L159

\bibitem[\protect\citeauthoryear{Igumenshchev}{2002}]{b14}
  Igumenshchev, I. V.,  2002, ApJ., 577, L31

\bibitem[\protect\citeauthoryear{Li, Ostriker \& Sunyaev}{2013}]{b15}
Li J., Ostriker J., Sunyaev R., 2013, ApJ., 767, 105

\bibitem[\protect\citeauthoryear{Meyer et al.}{2006a}]{b16}
Meyer L., Sch\"{o}del R., Eckart A., Karas V., Dov\v{c}iak M., Duschl W.J., 
2006a, A\&A, 458, L25

\bibitem[\protect\citeauthoryear{Meyer et al.}{2006b}]{b17}
Meyer L.,  Eckart A., Sch\"{o}del R.,  Duschl W.J., Mu\v{z}i\'{c} K., 
Dov\v{c}iak M., Karas, V., 2006b, A\&A, 460, 15



\bibitem[\protect\citeauthoryear{Mirabel et al.}{1998}]{b18}
Mirabel L. F,  Dhawan V., Chaty S., Rodr\'iguez L. F., Mart\'i J., Robinson C. R.,
Swank J., Geballe T. R., 1998, A\&A, 330, L9

\bibitem[\protect\citeauthoryear{Mo\'{s}cibrodzka, Das \& Czerny}{2006}]{b19}
Mo\'{s}cibrodzka M., Das T.K., Czerny B., 2006, MNRAS, 370, 219

\bibitem[\protect\citeauthoryear{Narayan, Igumenshchev \& Abramowicz}{2000}]{b20}
 Narayan R., Igumenshchev, I. V., Abramowicz, M. A., 2000, ApJ., 539, 798

\bibitem[\protect\citeauthoryear{Narayan et al.}{2012}]{b21}
  Narayan, R., S\"A dowski, A., Pen\'{n}a, RF., Kurkarni, AK., 2012, MNRAS, 426, 3241

\bibitem[\protect\citeauthoryear{Narayan \& Yi}{1994}]{b22}
 Narayan R., Yi I.,  1994, ApJ, 428, L13

\bibitem[\protect\citeauthoryear{Narayan \& Yi}{1995}]{b23}
 Narayan R., Yi I.,  1995, ApJ, 452, 710

\bibitem[\protect\citeauthoryear{Okuda}{2014}]{b24}
 Okuda T.,  2014, MNRAS, 441, 2354
 

 \bibitem[\protect\citeauthoryear{Okuda \& Molteni}{2012}]{b25}
 Okuda T.,  Molteni D., 2012, MNRAS, 425, 2413

\bibitem[\protect\citeauthoryear{Okuda, Fujita \& Sakashita}{1997}]{b26}
  Okuda T.,  Fujita M.,  Sakashita S.,  1997, PASJ, 49, 679

\bibitem[\protect\citeauthoryear{Okuda, Lipunova \& Molteni}{2009}]{b27}
  Okuda T.,  Lipunova G. V.,  Molteni D.., 2009, MNRAS, 398, 1668

\bibitem[\protect\citeauthoryear{Okuda, Teresi \& Molteni}{2009}]{b28}
  Okuda T.,  Teresi V.,  Molteni D.., 2007, MNRAS, 377, 1431

\bibitem[\protect\citeauthoryear{Quataert \& Gruzinov}{2000}]{b29}
 Quataert, E., Gruzinov, A., 2000, ApJ., 539, 809

 
 \bibitem[\protect\citeauthoryear{Shakura \& Sunyaev}{1973}]{b30}
 Shakura N.I.,   Sunyaev R.A., 1973, A\&A, 24, 337
 

 \bibitem[\protect\citeauthoryear{Shu}{1992}]{b31}
  Shu F. H., 1992,  Osterbrock D. E., Miller J. S., eds., The Physics of Astrophysics, 
 Volume I\hspace{-.1em}I: Gas Dynamics, University Science Books,  California
 
 
 

\bibitem[\protect\citeauthoryear{Stone \& Pringle}{2001}]{b32}
 Stone J. M.,  Pringle, J. E., 2001, MNRAS, 322, 461



\bibitem[\protect\citeauthoryear{Stone, Pringle \& Begelman}{1999}]{b33}
 Stone J. M., Pringle, J. E., Begelman, M. C., 1999, MNRAS, 310, 1002


\bibitem[\protect\citeauthoryear{Stepney \& Guilbert}{1983}]{b34}
 Stepney S.,  Guilbert P.W., 1983, MNRAS, 204, 1269

\bibitem[\protect\citeauthoryear{Trippe et al.}{2007}]{b35}
 Trippe S., Paumard T., Ott T., Gillessen S., Eisenhauer F., Martins F.,
Genzel, R., 2007, MNRAS, 375, 764



\bibitem[\protect\citeauthoryear{Yuan}{2011}]{b36}
Yuan F., 2011, in Morris M.R., Wang  Q.D., Yuan F., eds, 
ASP Conf. Ser. 439,
The Galactic Center: A Window to the Nuclear Environment of Disk 
 Galaxies. Astron. Soc. Pac., San Francisco, p. 346

\bibitem[\protect\citeauthoryear{Yuan, Bu \& Wu}{2012}]{b37}
 Yuan, F., Bu, D., Wu, M.. 2012, ApJ., 761, 130


\bibitem[\protect\citeauthoryear{Yuan et al.}{2015}]{b38}
Yuan F., Gan, Z., Narayan, R., S\"A dowski, A., Bu, D., Bai, XN,  2015, ApJ., 804, 101



\bibitem[\protect\citeauthoryear{Yuan et al.}{2009}]{b39}
 Yuan, F., Lin, J., Wu, K., Ho, L. C., 2009, MNRAS, 395, 2183

\bibitem[\protect\citeauthoryear{Yuan \& Narayan}{2014}]{b40}
Yuan F., Narayan R.,  2014,  ARA\&A, 52, 529


\bibitem[\protect\citeauthoryear{Yuan, Quataert \& Narayan}{2003}]{b41}
Yuan F., Quataert E., Narayan R.,  2003, ApJ, 598, 301


\bibitem[\protect\citeauthoryear{Yuan, Quataert \& Narayan}{2004}]{b42}
Yuan F., Quataert E., Narayan R.,  2004, ApJ, 606, 894



\bibitem[\protect\citeauthoryear{Yuan, Wu \& Bu}{2012}]{b43}
Yuan F., Wu M., Bu D.,  2012, ApJ, 761, 129



\bibitem[\protect\citeauthoryear{Yusef-Zadeh et al.}{2009}]{b44}
Yusef-Zadeh F. et al., 2009, ApJ, 706, 348

\bibitem[\protect\citeauthoryear{Yusef-Zadeh et al.}{2011}]{b45}
Yusef-Zadeh F., Wardle M., Miller-Jones J.C.A.,  Roberts D.A., 
Grosso N., Porquet D., 2011, ApJ, 729, 44


\end{thebibliography}
\end{document}